\documentclass[12pt,a4paper,oneside]{article}
\usepackage[top=20truemm,bottom=20truemm,left=20truemm,right=20truemm]{geometry}

\usepackage{hyperref}
\usepackage{amsmath, amsthm}
\usepackage{amssymb}
\usepackage{multirow}
\usepackage{graphicx}
\newcommand{\Trp}{\mathrm{Tr}'}

\numberwithin{equation}{section}

\begin{document}
\begin{flushright}
KEK-TH 1853
\end{flushright}
\begin{center}
{\LARGE Construction of 4d SYM compactified on open Riemann surfaces by the superfield formalism}\\
\vspace{5mm}
{\large Koichi Nagasaki${}^1$}\\
\vspace{5mm}
{\small ${}^1$KEK Theory Center, High Energy Accelerator Research Organization (KEK)\\
Address: 1-1 Oho, Tsukuba, Ibaraki, 305-0801, Japan\\
\texttt{nagasa@post.kek.jp}}\\
\end{center}

%%%%%%Abstract
%%%we can add the boundary degrees of freedom by super field formalism. X in the previous work, X half space, sentence 2,3 remove, adding the dof is difficult
\begin{abstract}
By compactifying gauge theories on a lower dimensional manifold, we often find many interesting relationships between geometry and supersymmetric quantum field theories.
In this paper we consider conformal field theories obtained from twisted compactification on a Riemann surface with a boundary. 
Various kinds of supersymmetric boundary conditions are exchanged under S-duality. 
To consider these transformations one need to take into account boundary degrees of freedom.
So we study how these degrees of freedom can be added at the boundary of the Riemann surface.
For these the boundary fields to be added it is convenient to rewrite the theory by means of superfields.
Therefore, I show in this paper that the 4d SYM action can be surely expressed as 2d superfields.
\end{abstract}

%%%%%%
\section{Introduction and summary}
% background
By compactifying gauge theories on lower dimensional surfaces, many interesting relationships between geometry and supersymmetric quantum field theories have been found. The Alday-Gaiotto-Tachikawa correspondence \cite{Alday:2009aq, Wyllard:2009hg} is an example of such theories. 
Related works are found in \cite{Bershadsky:1995vm, Dimofte:2011ju, Bah:2011vv, Bah:2012dg, Klemm:1996bj, Cecotti:2011iy,Okazaki:2014sga}.
Especially, the reduction of 4-dimensional theories on closed Riemann surfaces has been studied in various ways \cite{Bershadsky:1995vm, Benini:2012cz, Benini:2013cda}. 
We can preserve the supersymmetry on curved space by twisting the theory \cite{Bershadsky:1995qy, Maldacena:2000mw, 2011arXiv1111.4234C}. 

Studying effects of introducing a boundary is also an interesting theme \cite{1995NuPhB.455..522M, Cardy:2004hm, Petkova:2000dv}.
The boundary conditions for preserving the supersymmetry have been studied in many works \cite{Gaiotto:2008sa, Gaiotto:2008sd, Gaiotto:2008ak}.

% motivation of this work
We are interested in boundary conditions which can preserve some of supersymmeries.
An interesting work is to find these boundary conditions and classify them as in \cite{Gaiotto:2008sa, Gaiotto:2008sd, Gaiotto:2008ak, Hashimoto:2014vpa, Hashimoto:2014nwa}.
These various boundary theories are expected to be related by S-duality.
For example, under the S-duality transformation the NS5-like boundary condition \eqref{Eq:NS5cond} is transformed into the D5-like boundary condition \cite{Gaiotto:2008ak}. 
It is important to consider degrees of freedom located on the boundary because the number of the degrees of freedom can be in general changed under the S-duality transformation. 
Then, in order to study the structure of S-duality, it is needed to add these degrees of freedom to the theory.

% goal
The introduction of these boundary degrees of freedom is done in a natural way in terms of superfields.
Then, we would like to describe the 4-dimensional Yang-Mills Lagrangian in terms of 2-dimensional superfields.
This is the main goal of this paper. 
A similar construction of the 4-dimensional Lagrangian is done in \cite{Erdmenger:2002ex} where they described the bulk 4-dimensional theory by 3-dimensional superfields.
By the supersymmetry transformation of Yang-Mills fields \eqref{Eq:YMgaugetransf}, we found the theory compactified on the Riemann surface does not have translation symmetry along the boundary ($x^2$ direction) as well as the perpendicular direction ($x^3$). 
Then, in our situation, the theory has 2-dimensional supersymmetry.
Therefore we use 2-dimensional superfields to express the Yang-Mills Lagrangian.
Now, we specifically consider the $\mathcal{N}=(2,2)$ case.
This supersymmetry is realized when one adds the boundary to the theory with $\mathcal{N}=(4,4)$ supersymmetry.

% review of the previous work
In the previous work \cite{Nagasaki:2014xya} we considered such compactified theories on the Riemann surface with a boundary. 
We introduced a geodesic boundary on the Riemann surface and showed that one of the boundary condition called the NS5-like boundary condition can be imposed at the boundary.
As a result the half of the supersymmetries, $\mathcal{N}=(0,1), (1,1), (2,2)$, out of $\mathcal{N}=(0,2), (2,2), (4,4)$ can be preserved when the Riemann surface has the boundary. 
We consider the 4-dimensional super Yang-Mills theory on the space $R^{1,1}\times \Sigma_\mathbf{g}$ where $R^{1,1}$ is the 2-dimensional Minkowski space with the metric $g_{mn}=(-1,+1)$ and $\Sigma_\mathbf{g}$ is the Riemann surface with genus $\mathbf{g}$, which has the boundary.
We take the coordinates $(x^{m},x^{i})=(x^0,x^1,x^2,x^3),$ so that $R^{1,1}$ is spanned by the first two coordinates $x^m = x^0, x^1$ and the Riemann surface $\Sigma_\mathbf{g}$ is spanned by the coordinates $x^i = x^2,x^3$.

% resulet
The main result of this paper is that we can express the 4d Yang-Mills Lagrangian in terms of 2d superfields so that the remaining supersymmetry is manifest.
Our theory has 2d $\mathcal{N}=(2,2)$ supersymmetry. 
This is the 1/2 BPS state obtained from the 2d $\mathcal{N}=(4,4)$ on the closed Riemann surface when we introduce the boundary.

The 4-dimensional Yang-Mills Lagrangian is written in superfield formalism in Section \ref{Sec:NAbelian}.
According to our results, the action is written by 2-dimensional $\mathcal{N}=(2,2)$ superfields in flat space $(x^0,x^1)$:
\begin{align}\label{Eq:YMAction}
S_\text{YM}=\int d^4x\: \mathcal{L}^{g}
		&= \int_{R^{1,1}\times \Sigma_\mathbf{g}} d^4x\: \left({\mathcal{L}^{g}}_{\Sigma} + {\mathcal{L}^{g}}_\text{K1} 
				+ {\mathcal{L}^{g}}_\text{K2} + {\mathcal{L}^{g}}_{W}
\right)\nonumber\\
		&= \int_{R^{1,1}} d^2x \hat{\Trp}\sqrt{g}\:\int d^4\theta \bigg[ -\overline{\Sigma}\Sigma 
			+ 2e^{-2V}\overline{\Phi}_1e^{2V}\Phi_1
			\bigg]\nonumber\\
		&\hspace{1cm}	+ \int_{R^{1,1}} d^2x\hat{\Trp}\int d^4\theta \sum_{i=2,3} \bigg\{e^{-2V}\left(\frac12\partial_i +\overline{\Phi}_i\right)e^{2V} + \Phi_i\bigg\}^2
				\nonumber\\
		&\hspace{1cm}	+ 2 \int_{R^{1,1}} d^2x \hat{\Trp} \bigg[ 
		 \int d^2\theta\: \left(\Phi_1(\partial_2 \Phi_3 - \partial_3 \Phi_2)
					-2[\Phi_2,\Phi_3] \right) + \text{  c.c.  }
				\bigg],
\end{align}
where $g$ is the determinant of the metric on the Riemann surface $\Sigma_\mathbf{g}$ spanned by $(x^2,x^3)$.
Now $\Sigma_\mathbf{g}$ is regarded as an internal space.
Each term of the above action is concretely calculated in Section \ref{Sec:NAbelian}.

% boundary degrees of freedom 
Expressing the Lagrangian in terms of 2-dimensional superfields, it is easy to add other fields localized on the boundary.
For example, we can introduce fundamental and anti-fundamental superfields $B^{\pm}$ localized on the boundary.
These fields live in 3-dimensional spacetime corresponding to the boundary of 4-dimensional Yang-Mills theories ($x^3=0$).
The gauge transformations of these fields are
\begin{align} 
B^{+} \rightarrow e^{i\Lambda}B^{+},\:\:
B^{-} \rightarrow e^{-i\Lambda}B^{-},\:\: 
\Lambda = \Lambda (\theta, x^0,x^1,x^2).
\end{align}
Coupling of these boundary degrees of freedom to the bulk fields is 
\begin{align}\label{Eq:BdryAction}
{S}_\text{bdry} = \int_{R^{1,1}\times S^1} d^3x \int d^4\theta \left(
	\overline{B}_{+} e^{2V} B_{+} + \overline{B}_{-} e^{-2V} B_{-}
	\right),
\end{align}
where the integral is defined on the 3-dimensional spacetime where fields $B^{\pm}$ live.
The total action is obtained as the sum of \eqref{Eq:YMAction} and \eqref{Eq:BdryAction}.
 
 % organization 
This paper is organized as follows. 
Section \ref{Sec:4dYM} describes the outline of super Yang-Mills theory. 
This theory is twisted for preserving the supersymmetry on curved spacetime.
This super Yang-Mills theory is reproduced in terms of superfields in the rest of the paper.
Section \ref{Sec:Superfields} introduces the vector and chiral multiplets in 2-dimensions. 
Section \ref{Sec:Abelian} treats a simpler case for practice.
We treat in this section the Abelian theory on flat spacetime where the metric on the compactified space is simply $g_{ij} =\delta_{ij}$. 
Based on this result, in Section \ref{Sec:NAbelian} we generalize the theory to non-Abelian fields defined on curved space where the metric on the Riemann surface $g_{ij} = e^{2h}\delta_{ij}$ is introduced.
In this section we find the theory by the superfield formalism is equivalent to the super Yang-Mills on curved space explained in Section \ref{Sec:4dYM}. 
And we close with discussion in Section \ref{Sec:Discussion}.

%%%%%%
\section{Yang-Mills theory in 4-dimension}\label{Sec:4dYM}
Our theory is constructed on the space $R^{1,1}\times \Sigma_\mathbf{g}$. 
The first factor $R^{1,1}$ is a flat space with metric $g_{mn}=\text{diag}(-1,+1)$ and the second factor is a Riemann surface with genus $\mathbf{g}$. 
We choose the coordinates in which the metric on the Riemann surface is 
\begin{align}
g_{ij} = e^{2h(x^2,x^3)}\delta_{ij},
\end{align} 
where $h(x^2,x^3)$ is a function of $x^2$ and $x^3$.
In the following we use the indices $m,n$ for the flat direction $x^0,x^1$, the indices $i,j$ for the Riemann surface with coordinates $x^2,x^3$ and the indices $\mu,\nu$ are used for the whole of them.

The gauge field is $A_\mu;\: \mu=0,1,2,3$ and the field strength is defined as
\begin{align}
F_{\mu\nu} = \partial_\mu A_\nu -\partial_\nu A_\mu + i[A_\mu, A_\nu]. 
\end{align}
There are 6 scalar fields, $X_A;\: A=4,\cdots 9$, and the fermion field $\Psi$ which is a Majorana-Weyl spinor satisfying, $-\Gamma^{0123456789}\Psi = \Psi$.
The 4-dimensional $\mathcal{N}=4$ super Yang-Mills Lagrangian is 
\begin{align}\label{Eq:YMlagrangian}
\mathcal{L}_\text{gauge} = \frac{1}{(g_\text{YM})^2}\sqrt{g}\:\text{Tr}'
		\Big\{
			-\frac{1}{4}F_{\mu\nu}F^{\mu\nu}
			-\frac{1}{2}\nabla_\mu X_A \nabla^\mu X^A
			+ \frac{1}{4}[X_A,X_B][X^A,X^B]\nonumber\\
			+\frac{i}{2}\overline{\Psi}\Gamma^\mu \nabla_\mu \Psi
			-\frac{1}{2}\overline{\Psi}\Gamma^A[X_A,\Psi]
			\Big\}.
\end{align}
$\Trp$ is  a trace normalized as $\Trp=\frac{1}{h^{\vee}} \mathrm{Tr}_{\text{adjoint}}$ where $h^{\vee}$ is the dual Coxeter number.
In the above the covariant derivative for the fields are defined as
\begin{subequations}\label{Eq:Covderiv}
\begin{align}
\nabla_\mu X_A &:= \partial_\mu X_A + i[A_\mu, X_A],\\
\nabla_\mu \Psi &:= \partial_\mu \Psi + i[A_\mu, \Psi].
\end{align}
\end{subequations}

By twisting this action, we obtain the Lagrangian which is invariant under the supersymmetry transformation:
\begin{align}\label{Eq:YMgaugetransf}
\delta A_\mu 	&= i\overline{\epsilon}\Gamma_\mu \Psi,\:\:
\delta X_A' 	  = i\overline{\epsilon}\Gamma_A \Psi,\nonumber\\
\delta \Psi 		&= \frac{1}{2}\Gamma^{\mu\nu}F_{\mu\nu}\epsilon 
				+ \frac{i}{2}\Gamma^{AB}[X_A',X_B']\epsilon
				+ \Gamma^{\mu A}\nabla_\mu' X_A'\epsilon,
\end{align}
where $\prime$ denotes the twisted fields.

Our goal is to describe the above Lagrangian in terms of superfields.
Similar construction of the 4-dimensional Lagrangian is done in the paper \cite{Erdmenger:2002ex} where they described the bulk 4-dimensional theory by 3-dimensional superfields.

In order to preserve the supersymmetries on such a curved space, we usually need to twist the theory. 
According to the previous work \cite{Nagasaki:2014xya}, we know $\mathcal{N}=(2,2)$ supersymmetry is obtained from the no boundary with $\mathcal{N}=(4,4)$ case. 
These supersymmetries are generated by the parameter satisfying,
\begin{align}\label{Eq:EpsilonCond}
\Gamma^{2345}\epsilon = -\epsilon,\:
\Gamma^{3579}\epsilon = +\epsilon.
\end{align}
These conditions can be written as
\begin{align}
P_{1+} \epsilon = 0,\:\:  P_{1-} \epsilon = \epsilon,\\
P_{2+} \epsilon = \epsilon,\:\:  P_{2-} \epsilon = 0,
\end{align}
where the projection operators are defined as follows:
\begin{align}
P_{1\pm} := \frac{1\pm\Gamma^{2345}}{2},\:
P_{2\pm} := \frac{1\pm\Gamma^{3579}}{2}.
\end{align}

In our coordinates the boundary corresponds to $x^3=0$.
One of the boundary condition which preserves $\mathcal{N}=(2,2)$ supersymmetries is the NS5-like boundary condition defined as
\begin{subequations}\label{Eq:NS5cond}
\begin{align}
F_{\mu 3}|	_\text{bdry}	&=0\: (\mu=0,1,2),\\ 
D_3X_A 	|_\text{bdry}	&=0\; (A=4,5,6),\\
X_A		|_\text{bdry}	&=0\; (A=7,8,9).
\end{align}
\end{subequations}

%%%%%%%%
\section{Superfields in 2-dimension}\label{Sec:Superfields}
In order to construct the 4-dimensional Yang-Mills Lagrangian \eqref{Eq:YMlagrangian} in terms of superfields, we introduce a vector multiplet and three chiral multiplets.
As we noted in Introduction, our theory has Poincare symmetry in 2-dimensional space $(x^0,x^1)$.
Then, we introduce the 2-dimensional superfields.
We use the notation used in \cite{Witten:1993yc, Witten:1993xi}.

In the superspace coordinates,$(x^0,x^1,\theta^{\pm},\bar\theta^{\pm})$, differential operators of the superspace are shown in Appendix \ref{Sec:Notation}. 
\begin{align}\label{Eq:differentialOp}
D_\pm 	:= \frac{\partial}{\partial \theta^{\pm}} - 2i\bar{\theta}^\pm\partial_\pm,\:\:
\overline{D}_\pm := -\frac{\partial}{\partial \bar{\theta}^\pm} + 2i\theta^\pm\partial_\pm.
\end{align}
In the above expression we used $x^{\pm}:= x^0 \pm x^1$, $\partial_\pm = \frac{1}{2}(\partial_0 \pm \partial_1)$. 
For spinors the indices $\pm$ are raised or lowered by the epsilon tensor $\epsilon^{ij}$; $\epsilon^{-+}=+1 =- \epsilon_{-+}$.

We use the notation for integration for Grassmann coordinates:
\begin{align}
\int d^4\theta\: F &= F \Big|_{\theta\theta\bar\theta\bar\theta} 
				= \frac{1}{4} F\Big|_{\theta^{-}\theta^{+}\bar\theta^{+}\bar\theta^{-}},\\
\int d^2\theta\: F &= F \Big|_{\theta\theta} 
				= \frac{1}{2} F\Big|_{\theta^{-}\theta^{+}}.
\end{align}
We use an unusual ordering of sigma matrices:
\begin{align}
\sigma^\mu = \left\{
	\sigma^{m}, \sigma^{i}
	\right\}
	= \left\{
\left(\begin{array}{cc}-1 & 0 \\0 & -1\end{array}\right),
\left(\begin{array}{cc}1 & 0 \\0 & -1\end{array}\right),
\left(\begin{array}{cc}0 & 1 \\1 & 0\end{array}\right),
\left(\begin{array}{cc}0 & -i \\i & 0\end{array}\right)
\right\}.
\end{align}

%%%%%
\subsection{Vector multiplet}
We choose the Wess-Zumino gauge.
The 2-dimensional vector multiplet is
\begin{align}\label{Eq:VecFields}
V 	= -\sum_{m=0,1}\theta\sigma^m \bar\theta v_m (x)
	 -\sum_{a=2,3}\theta\sigma^a \bar\theta v_a (x)
	 + i\theta\theta\bar\theta\bar\lambda(x)
	 - i\bar\theta\bar\theta\theta\lambda(x)
	 + \frac{1}{2}\theta\theta\bar\theta\bar\theta D(x),
\end{align}
where $v_m,\: m=0,1$, are the component of the 2-dimensional vector field, $v_a,\: a=2,3$, are scalar fields, $\lambda$ and $\overline{\lambda}$ are fermion fields and $D$ is an auxiliary field.
%We renamed sigma matrices as $(\sigma^2, \sigma^3) = (\sigma^6, \sigma^8)$.
We also express the vector multiplet by components $\theta = (\theta^{-}, \theta^{+})$,
\begin{align}\label{Eq:VecComp}
V &= 2\theta^{-}\bar{\theta}^{-}v_{-} + 2\theta^{+}\bar{\theta}^{+}v_{+}
	-\sqrt{2}\bar{\sigma}\theta^{-}\bar\theta^{+} - \sqrt{2}\sigma\theta^{+}\bar{\theta}^{-}\nonumber\\
	&\qquad
	-2i\theta^{-}\theta^{+}(\bar{\theta}^{-}\overline{\lambda}_{-} + \bar{\theta}^{+}\overline{\lambda}_{+})
	-2i\bar\theta^{+}\bar\theta^{-}(\theta^{-}\lambda_{-} + \theta^{+}\lambda_{+})
	+2\theta^{-}\theta^{+}\bar\theta^{+}\bar\theta^{-}D.
\end{align}
We redefined the vector and scalar fields as follows:
\begin{align}
2v_{\pm} := v_{0} \pm v_{1},\:
\sqrt{2}\sigma := v_2 -iv_3.
\end{align}
%The gauge transformation of this field is expressed in components as:
%\begin{subequations}\label{Eq:VecGaugetransf}
%\begin{align}
%\delta v_m 	&= i(\xi\sigma_m\overline{\lambda} - \lambda\sigma_m\overline{\xi}),\\
%\delta v_a 		&= i(\xi\sigma_a\overline{\lambda} - \lambda\sigma_a\overline{\xi}),\\
%\delta \lambda_\alpha &= i\xi_\alpha D 
%	+ \left(\frac{1}{4}(\sigma^{\mu}\bar{\sigma}^{\nu} - \sigma^{\nu}\bar{\sigma}^{\mu})\xi\right)_{\alpha} v_{\mu\nu}.
%\end{align}
%\end{subequations}

%%%%%
\subsection{Chiral multiplets}
Our theory has the three chiral multiplets $\Phi_i$ for $i=1,2,3$:
\begin{align}\label{Eq:ChiralFields}
\Phi_i = \phi_i + \sqrt{2}\theta\psi_i +i\theta\sigma^m\bar\theta \partial_m \phi_i
		+ \theta\theta F_i 
		+ \frac{i}{\sqrt{2}}\theta\theta\bar\theta\bar{\sigma}^m\partial_m\psi_i 
		+ \frac{1}{4}\theta\theta\bar\theta\bar\theta \square \phi_i,
\end{align}
where $\phi_i$ are bosonic, $\psi_i$ are fermonic, and $F_i$ is a bosonic auxiliary field.
We also use their harmitian conjugate are 
\begin{align}\label{Eq:ChiralFieldsh}
\overline{\Phi}_i = \overline{\phi}_i + \sqrt{2}\bar{\theta}\overline{\psi}_i 
		-i\theta\sigma^m\bar\theta \partial_m \overline{\phi}_i
		+ \bar{\theta}\bar{\theta} \overline{F}_i 
		+ \frac{i}{\sqrt{2}}\bar{\theta}\bar{\theta}\theta\sigma^m\partial_m\overline{\psi}_i 
		+ \frac{1}{4}\theta\theta\bar\theta\bar\theta \square \overline{\phi}_i.
\end{align}
%The gauge transformation of the chiral multiplets is
%\begin{subequations}\label{Eq:ChiralGaugetransf} 
%\begin{align}
%\delta \phi_i 	&= \sqrt{2}\xi\psi_i,\\
%\delta (\psi_i)_{\alpha} &= \sqrt{2}\xi_{\alpha}F_i + \sqrt{2}i (\sigma^m\overline{\xi})_{\alpha}\partial_m\phi_i.
%\end{align}
%\end{subequations}

%%%%%
\subsection{Correspondence to the 4d fields}
Getting together the transformations \eqref{Eq:VecGaugetransf} and \eqref{Eq:ChiralGaugetransf} we obtain the 4-dimensional gauge transformation \eqref{Eq:YMgaugetransf} with the parameter $\epsilon = (\xi, \overline\xi)$.
The degrees of freedom surely agrees between 4-dimensional fields, $A_\mu$ and $\Psi$, and one vector multiplet and three chiral multiplets.
We summarize the relation between the chiral and vector multiplets and the 4-dimensional fields in Table \ref{Table:CorrespondenceFields}. 
This correspondence can be shown by using the property of the supersymmetry parameter $\epsilon$ \eqref{Eq:EpsilonCond} and is listed in the right side of Table \ref{Table:CorrespondenceFields}. 

\begin{table}[b]
\centering
  \begin{tabular}{| l  c  l | l |}
    \hline
    2d fields 		&  				& 4d fields & relation\\ \hline
    $v_0$, $v_1$ 	& $\longleftrightarrow$ & $A_0$, $A_1$ & $v_0 = A_0,\:\: v_1 = A_1$ \\ \hline
    $v_2$, $v_3$ 	& $\longleftrightarrow$ & $X_6$, $X_8$ & $v_2 = X_6,\:\: v_3 = X_8$ \\ \hline
    $\:\: \lambda$ 	& $\longleftrightarrow$ & $P_{1-}P_{2+}\Psi$ &  \\ \hline
    $\:\: \phi_1$	& $\longleftrightarrow$ & $X_7$, $X_9$ & $\phi_1 = \frac{1}{2}X_7 + i\frac{1}{2}X_9$ \\ \hline
    $\:\: \psi_1$  & $\longleftrightarrow$ & $P_{1-}P_{2-}\Psi$ & \\ \hline
    $\:\: \phi_2$ 	& $\longleftrightarrow$ & $A_2$, $X_4$ & $\phi_2 = \frac{1}{2}X_4 - i\frac{1}{2}A_2$ \\ \hline
    $\:\: \psi_2$ 	& $\longleftrightarrow$ & $P_{1+}P_{2+}\Psi$ & \\ \hline
    $\:\: \phi_3$ 	& $\longleftrightarrow$ & $A_3$, $X_5$ & $\phi_3 = \frac{1}{2}X_5 - i\frac{1}{2}A_3$ \\ \hline
    $\:\: \psi_3$ 	& $\longleftrightarrow$ & $P_{1+}P_{2-}\Psi$ & \\ \hline
  \end{tabular}
\caption{Component correspondence (Vector and chiral multiplets)}
\label{Table:CorrespondenceFields}
\end{table}

%%%%%%%%%
\section{Simple case}\label{Sec:Abelian}
Our goal is to construct the Lagrangian \eqref{Eq:YMlagrangian} in terms of superfields \eqref{Eq:VecFields} and \eqref{Eq:ChiralFields}.
First, in this section we consider a simple case where the fields are Abelian and the metric is simply flat $g_{ij} = \eta_{ij}$. 
In the next section we consider a non-Abelian case and introduce the curved metric.
The total action in terms of the superfields is
\begin{align}\label{}
\int d^4x\: \mathcal{L} 
		&= \int d^4x\: \left(\mathcal{L}_{\Sigma} + \mathcal{L}_\text{K1} + \mathcal{L}_\text{K2} + \mathcal{L}_{W}
\right)\nonumber\\
		&= \int d^4x \bigg\{- \int d^4\theta\: \overline{\Sigma}\Sigma 
			+ 2 \int d^4\theta\: \overline{\Phi}_1\Phi_1 + \int d^4\theta \sum_{i=2,3} (\partial_iV+\overline{\Phi}_i + \Phi_i)^2 
				\nonumber\\
		&\hspace{2cm}	+ 2 \left(\int d^2\theta\: \Phi_1(\partial_2 \Phi_3 - \partial_3 \Phi_2)
				+ \text{  c.c.  }
				\right)\bigg\}.
\end{align}
In the above expression the normalization of each term is defined so that this Lagrangian gives the 4-dimensional Lagrangian \eqref{Eq:YMlagrangian}.

We construct the each term in the following subsections and give the bosonic part of the Lagrangian:
\begin{align}\label{Eq:Lagrangian}
\mathcal{L}_\text{bos} = -\frac14 F_{\mu\nu} F^{\mu\nu} -\frac12 \sum_{A=4,\cdots,9}\partial_\mu X_A\partial^\mu X_A.
\end{align}
In the above $F_{\mu\nu}$ is the field strength of the Abelian gauge field,
\begin{equation}
F_{\mu\nu} := \partial_\mu A_\nu - \partial_\nu A_\mu.
\end{equation}

%%%%%
\subsection{Kinetic term of the vector multiplet: $\mathcal{L}_{\Sigma}$}
This term is constructed by the twisted chiral superfield defined by the vector superfied \eqref{Eq:VecComp} as 
\begin{align}
\Sigma := \frac{1}{\sqrt{2}}\overline{D}_{+}D_{-}V.
\end{align}
Substituting the expression \eqref{Eq:differentialOp}, we calculate 
\begin{align}
\overline{\Sigma}\Sigma\Big|_{\theta\theta\bar\theta\bar\theta}
	&=\frac{1}{4}\overline{\Sigma}\Sigma\Big|_{\theta^{-}\theta^{+}\bar\theta^{+}\bar\theta^{-}}\nonumber\\
	&= -\frac{1}{2}D^2 - \frac{1}{2}(2v_{-+})^2 
		- 2i( \bar\lambda_{-}\partial_{+}\lambda_{-}
			+ \bar\lambda_{+}\partial_{-}\lambda_{+})
		+4\bar\sigma\partial_{+}\partial_{-}\sigma.
\end{align}

The Lagrangian of the linear multiplet part is
\begin{align}\label{Eq:LagSigma}
\mathcal{L}_{\Sigma} 
		&= -\overline{\Sigma}\Sigma\Big|_{\theta\theta\bar\theta\bar\theta}\nonumber\\
		&= \frac{1}{2}D^2 + \frac{1}{2}v_{01}^2
					+ 2i (\overline{\lambda}_{-}\partial_{+}\lambda_{-} + \overline{\lambda}_{+}\partial_{-}\lambda_{+})
					+ \bar\sigma( -\partial_0^2 + \partial_1^2)\sigma.
\end{align}
In the above to obtain the last expression we ignore the total derivative terms.
%%%%%
\subsection{Kinetic term of the chiral multiplet, $\Phi_1$: $\mathcal{L}_\text{K1}$}
The kinetic term of the field $\Phi_1$ is obtained in the usual way:
\begin{align}
\mathcal{L}_\text{K1} = 2\int d^4\theta \overline{\Phi}_1\Phi_1.
\end{align}
Substituting the component expansion \eqref{Eq:ChiralFields} and \eqref{Eq:ChiralFieldsh}, we obtain the kinetic Lagrangian of the field $\Phi_1$:
\begin{align}\label{Eq:LagK1}
\mathcal{L}_\text{K1} 
	&= \overline{F}_1 F_1 + \overline\phi_1\square\phi_1 - i\overline\psi_1\bar\sigma^{m} \partial_{m}\psi_1.
\end{align}

%%%%%
\subsection{Kinetic term of the chiral multiplets, $\Phi_2, \Phi_3$:  $\mathcal{L}_\text{K2}$}
The kinetic term including the fields $\Phi_2,\Phi_3$ is 
\begin{align}
\mathcal{L}_\text{K2} = 
		\int d^4\theta \left(
		(\partial_2V + \overline\Phi_2 + \Phi_2)^2
		+(\partial_3V + \overline\Phi_3 + \Phi_3)^2
		\right).
\end{align}
We obtain the kinetic term of the chiral multiplet $\Phi_2$ and $\Phi_3$ up to total derivative terms:
\begin{align}\label{Eq:LagK2}
&\int d^4\theta \left(
		(\partial_2V + \overline\Phi_2 + \Phi_2)^2
		+(\partial_3V + \overline\Phi_3 + \Phi_3)^2
		\right)\nonumber\\
&	= -2D (\partial_2\text{Re}\phi_2 + \partial_3\text{Re}\phi_3) 
		+ 2(\overline{F}_2F_2 + \overline{F}_3F_3)
		+ 2(\text{Re}\phi_2 \square \text{Re}\phi_2 + \text{Re}\phi_3 \square \text{Re}\phi_3)\nonumber\\
&\qquad		+ i\sqrt{2}\psi_2\left(\partial_2\lambda - \frac{1}{\sqrt{2}}\sigma^m\partial_m\overline{\psi}_2\right)
		- i\sqrt{2}\overline\psi_2\left(\partial_2\overline\lambda + \frac{1}{\sqrt{2}}\bar\sigma^m\partial_m\psi_2\right)\nonumber\\
&\qquad		+ i\sqrt{2}\psi_3\left(\partial_3\lambda - \frac{1}{\sqrt{2}}\sigma^m\partial_m\overline{\psi}_3\right)
		- i\sqrt{2}\overline\psi_3\left(\partial_3\overline\lambda + \frac{1}{\sqrt{2}}\bar\sigma^m\partial_m\psi_3\right)\nonumber\\
&\qquad		-\frac{1}{2}(\partial_2v^m + 2\partial^m\text{Im}\phi_2)(\partial_2v_m + 2\partial_m\text{Im}\phi_2)
			-\frac{1}{2}(\partial_3v^m + 2\partial^m\text{Im}\phi_3)(\partial_3v_m + 2\partial_m\text{Im}\phi_3)
			\nonumber\\
&\qquad		-\frac{1}{2}\partial_2v^a\partial_2v_a -\frac{1}{2}\partial_3v^a\partial_3v_a.			
\end{align}

%%%%%
\subsection{Potential term: $\mathcal{L}_{W}$}
The potential term is constructed as 
\begin{align}
\mathcal{L}_{W} = 2 \left(\int d^2\theta\: \Phi_1(\partial_2 \Phi_3 - \partial_3 \Phi_2) + \text{  c.c.  }
				\right).
\end{align}
In the above expression, ``c.c." means complex conjugate of the first term.
In the component expression \eqref{Eq:ChiralFields} and \eqref{Eq:ChiralFieldsh}, this term becomes
\begin{align}\label{Eq:LagW}
\int d^2\theta \Phi_1(\partial_2\Phi_3 - \partial_3\Phi_2)
	&= \phi_1(\partial_2F_3 - \partial_3F_2) 
		+ F_3(\partial_2\phi_3 - \partial_3\phi_2) 
		- \psi_3 (\partial_2\psi_3 - \partial_3\psi_2)\nonumber\\
	&= F_1 (\partial_2F_3 - \partial_3F_2) + F_2\partial_3\phi_1 - F_3\partial_2\phi_1 - \psi_1(\partial_2\psi_3 - \partial_3\psi_2).
\end{align}
In the above, to obtain the last expression we ignore total derivative terms.

%%%%%
\subsection{Total Lagrangian $\mathcal{L}$}
Putting together the kinetic terms of the vector and chiral multiplets \eqref{Eq:LagSigma}, \eqref{Eq:LagK1}, \eqref{Eq:LagK2} and the potential term \eqref{Eq:LagW} we obtain the total Lagrangian:
\begin{align}
\mathcal{L}_{} &:= \mathcal{L}_{\Sigma} + \mathcal{L}_\text{K1} + \mathcal{L}_\text{K2} + \mathcal{L}_{W}\nonumber\\
	&= 	\frac{1}{2}D^2 + \frac{1}{2}v_{01}^2
					+ 2i (\overline{\lambda}_{-}\partial_{+}\lambda_{-} + \overline{\lambda}_{+}\partial_{-}\lambda_{+})
					+ \bar\sigma( -\partial_0^2 + \partial_1^2)\sigma\nonumber\\
	&\qquad	+\overline{F}_1 F_1 + \overline\phi_1\square\phi_1 - i\overline\psi_1\bar\sigma^{m} \partial_{m}\psi_1\nonumber\\
	&\qquad	+2\text{Re}\phi_2 (\partial_2 D + \square \text{Re}\phi_2)\nonumber\\
	&\qquad + i\sqrt{2}\psi_2 \left(\partial_2\lambda - \frac{1}{\sqrt{2}}\sigma^{m}\partial_{m}\overline\psi_2
						\right)
			- i\sqrt{2}\overline{\psi}_2 \left(\partial_2\overline\lambda + \frac{1}{\sqrt{2}}\bar\sigma^{m}\partial_{m}\psi_1
						\right)\nonumber\\
	&\qquad + 2\overline{F}_2 F_2 
		-\frac{1}{2}(\partial_2v^m + 2\partial^m \text{Im}\phi_2)
				(\partial_2v_m + 2\partial_m \text{Im}\phi_2)
		-\frac{1}{2}\partial_2 v^a \partial_2 v_a\nonumber\\
	&\qquad	+F_1 (\partial_2F_3 - \partial_3F_2) + F_2\partial_3\phi_1 - F_3\partial_2\phi_1 - \psi_1(\partial_2\psi_3 - \partial_3\psi_2).
\end{align}

We obtain the equations of motion of auxiliary fields: 
\begin{subequations}
\begin{align}
D :	&\: D - 2 (\partial_2\text{Re}\phi_2 + \partial_3\text{Re}\phi_3) =0,\\
F_1: &\: \overline{F}_1 + (\partial_2\phi_3 - \partial_3\phi_2) =0,\\
F_2: &\: \overline{F}_2 + \partial_3\phi_1 =0, \\
F_3: &\: \overline{F}_3 - \partial_2\phi_1 =0.
\end{align}
\end{subequations}
Eliminating these auxiliary fields and rewrite this Lagrangian in terms of 4-dimensional fields according to the correspondence (see Table \ref{Table:CorrespondenceFields}), we obtain the Lagrangian \eqref{Eq:Lagrangian}.

%%%%%%%%%
\section{Non-Abelian theory on curved space}\label{Sec:NAbelian}
In this section we would like to generalize the action in the previous theory into the non-Abelian on curved space. 
We consider the theory on Riemann surfaces with genus $\mathbf{g}$ whose metric is written as
\begin{align}
g_{ij} = e^{2h(x^2,x^3)}\delta_{ij}.
\end{align} 
The spin connection and the curvature are calculated as
\begin{align}
\Omega^{23} = \partial_3h dx^2 - \partial_2h dx^3,\\
\sqrt{g}R = -2\sum_{i}\partial_i\partial_ih\label{Eq:Curvature}.
\end{align}
In the case of theories on curved space, we need to twist the theory in order to preserve the supersymmetry.

The total action in terms of the superfields is
\begin{align}\label{Eq:NALagrangian}
\int d^4x\: \mathcal{L}^{g}
		&= \int d^4x\: \left({\mathcal{L}^{g}}_{\Sigma} + {\mathcal{L}^{g}}_\text{K1} 
				+ {\mathcal{L}^{g}}_\text{K2} + {\mathcal{L}^{g}}_{W}
\right)\nonumber\\
		&= \int d^4x \sqrt{g}\:{\Trp} \bigg[ - \int d^4\theta\: \overline{\Sigma}\Sigma 
			+ 2 \int d^4\theta\: e^{-2V}\overline{\Phi}_1e^{2V}\Phi_1
			\bigg]\nonumber\\
		&\hspace{1cm}	+ \int d^4x{\Trp}\int d^4\theta \sum_{i=2,3} \bigg\{e^{-2V}\left(\frac12\partial_i +\overline{\Phi}_i\right)e^{2V} + \Phi_i\bigg\}^2
				\nonumber\\
		&\hspace{1cm}	+ 2\int d^4x {\Trp} \bigg[ 
		 \int d^2\theta\: \left(\Phi_1(\partial_2 \Phi_3 - \partial_3 \Phi_2)
					-2[\Phi_2,\Phi_3] \right) + \text{  c.c.  }
				\bigg].
\end{align}
We construct the each term in the following subsections and show this Lagrangian gives the -Mills theory:
\begin{align}\label{Eq:NLagrangian}
{\mathcal{L}^{g}}_\text{bos} 
	= \sqrt{g}\:\Trp\Bigg[
	-\frac14 F_{\mu\nu} F^{\mu\nu} -\frac12 \sum_{A=4,\cdots,9}\nabla'_{\mu} X'_A \nabla'^{\mu} X'_A 
	+ \frac14 \sum_{A,B=4,\cdots, 9} [X'_A, X'_B]^2
	 -\frac14 R\sum_{A=4,5}(X'_AX'_A)\Bigg],
\end{align}
where the covariant derivative is defined as
\begin{align}
\nabla_{\mu} X'_A &:= \partial_{\mu}X'_A + i[A_{\mu}, X'_A] + \sum_{B}\mathcal{A}^{AB}_iX'_B,
\end{align}
and $X'_A;\: A=4,\cdots, 9$ are scalar fields on curved space.

In the last part of this section, \ref{Subsec:TotalNonabelLag}, we see that the superfield Lagrangian\eqref{Eq:NALagrangian} actually produces the 4-dimensional Yang-Mills Lagrangian \eqref{Eq:YMlagrangian}.

%%%%%
\subsection{Kinetic term of the vector multiplet: ${\mathcal{L}^{g}}_{\Sigma}$}
In this subsection we calculate the kinetic term of the vector field,
\begin{equation}
{\mathcal{L}^{g}}_{\Sigma} = - \int d^4\theta \sqrt{g}\:\Trp\: \overline{\Sigma}\Sigma.
\end{equation}

Let us define the differential operators ($\alpha = \pm$)                                                    
\begin{align}\label{Eq:NAbelianDiffOp}
\mathcal{D}_{\alpha}
	:= e^{-V} D_\alpha e^{V},\\
\overline{\mathcal{D}}_{\alpha}
	:= e^{V} \overline{D}_\alpha e^{-V}.
\end{align}

The superfield strength is defined as 
\begin{align}\label{Eq:LinearMulti}
\Sigma := \frac{1}{2\sqrt{2}}\{\overline{\mathcal{D}}_{+}, \mathcal{D}_{-}\}.
\end{align}

The differential operators in \eqref{Eq:LinearMulti} are calculated by the component expression of the vector super field \eqref{Eq:VecComp}:
\begin{align}
\overline{\mathcal{D}}_{+} 
	&= -\frac{\partial}{\partial\bar\theta^{+}} + \sqrt{2}\theta^{-}\sigma 
		+ 2i\theta^{+}(\partial_{+} + iv_{+})
		- 2i \theta^{-}\theta^{+}\bar\lambda_{+}
		- 2i \bar\theta^{-}(\theta^{-}\lambda_{-} + \theta^{+}\lambda_{+})\nonumber\\
	&\qquad	+ 2\theta^{-}\theta^{+}\bar\theta^{-} 
			\left( D+ 2i \partial_{+}v_{-} + [v_{-},v_{+}] + \frac{1}{2}[\sigma,\bar\sigma]
				\right)\nonumber\\
	&\qquad	- 2\sqrt{2}i \theta^{-}\theta^{+}\bar\theta^{+}\nabla_{+}\sigma
		+ 4\theta^{-}\theta^{+}\bar\theta^{+}\bar\theta^{-} 
			\left( \nabla_{+}\lambda_{-} + \frac{i}{2}[\sigma,\lambda_{+}]
			\right),\\
\mathcal{D}_{-} 
	&= \frac{\partial}{\partial\theta^{-}} - \sqrt{2}\bar\theta^{+}\sigma 
		- 2i\bar\theta^{-}(\partial_{-} + iv_{-})
		- 2i \bar\theta^{+}\bar\theta^{-}\lambda_{-}
		- 2i \theta^{+}(\bar\theta^{+}\bar\lambda_{+} + \bar\theta^{-}\bar\lambda_{-})\nonumber\\
	&\qquad	- 2\theta^{+}\bar\theta^{+}\bar\theta^{-} 
			\left( -D+ 2i \partial_{-}v_{+} - [v_{-},v_{+}] + \frac{1}{2}[\sigma,\bar\sigma]
				\right)\nonumber\\
	&\qquad	+ 2\sqrt{2}i \theta^{-}\bar\theta^{+}\bar\theta^{-}\nabla_{-}\sigma
		+ 4\theta^{-}\theta^{+}\bar\theta^{+}\bar\theta^{-} 
			 \left( \nabla_{-}\bar\lambda_{+} + \frac{i}{2}[\sigma,\bar\lambda_{-}]
			\right).
\end{align}
Then, we obtain the superfield strength in components 
\begin{align}
\Sigma &= \sigma + \sqrt{2}i(\bar\theta^{-}\lambda_{-} - \theta^{+}\overline{\lambda}_{+})
		+ 2i(\theta^{-}\bar\theta^{-}\nabla_{-}\sigma - \theta^{+}\bar\theta^{+}\nabla_{+}\sigma)
		+ \sqrt{2}\theta^{+}\bar\theta^{-}( D - iv_{01})\nonumber\\
	&\qquad	- \sqrt{2}\theta^{-}\theta^{+}\bar\theta^{-}
				\left( 2\nabla_{-}\overline{\lambda}_{+} + \sqrt{2}i[\sigma, \overline{\lambda}_{-}]
				\right)\nonumber\\
	&\qquad	+ \sqrt{2}\theta^{+}\bar\theta^{+}\bar\theta^{-}
				\left( 2\nabla_{+}\lambda_{-} + \sqrt{2}i[\sigma,\lambda_{+}]
					\right)\nonumber\\
	&\qquad	+ \theta^{-}\theta^{+}\bar\theta^{+}\bar\theta^{-}
			\left((\nabla_0^2 - \nabla_1^2)\sigma + i[\sigma,\partial_mv^m] - [\sigma,[\sigma,\bar\sigma]]
			\right).
\end{align}
We defined the field strength 
\begin{equation}
v_{-+} := -i[\partial_{-} -iv_{-}, \partial_{+} -iv_{+}]
\end{equation}
and write it in the coordinates $x_0$ and $x_1$,
\begin{equation}
v_{-+} = v_{01}.
\end{equation}
The linear multiplet term of the Lagrangian is
\begin{align}
{\mathcal{L}^{g}}_{\Sigma} &= -\int d^4\theta\sqrt{g}\:{\Trp}\: \overline\Sigma\Sigma\nonumber\\
	&= \sqrt{g}\Big\{ \frac{1}{2}(D^2+v_{01}^2) 
		+ |\nabla_0\sigma |^2 - |\nabla_1\sigma |^2 - \frac{1}{2}[\sigma,\bar\sigma]^2\nonumber\\
	&\qquad	+ 2i(\overline{\lambda}_{-}\nabla_{+}\lambda_{-} + \overline{\lambda}_{+}\nabla_{-}\lambda_{+})
		 -\sqrt{2}(\lambda_{+}[\sigma,\overline{\lambda}_{-}] + \overline{\lambda}_{+}[\bar\sigma,\lambda_{-}])\Big\}.
\end{align}

%%%%%
\subsection{Kinetic term of the chiral multiplet, $\Phi_1$: ${\mathcal{L}^{g}}_\text{K1}$}
Using the differential operators \eqref{Eq:NAbelianDiffOp}, the chiral superfield is given by the condition  
\begin{equation}
\overline{{D}}_{\pm}\Phi_i = 0.
\end{equation}

The kinetic term of the chiral multiplets $\Phi_1$ is defined as
\begin{align}
{\mathcal{L}^{g}}_\text{K1}
	= 2\int d^4\theta\sqrt{g}\:{\Trp}\left(
		e^{-2V}\overline{\Phi}_1 e^{2V}\Phi_1
		\right).
\end{align}
Substituting the component expression \eqref{Eq:ChiralFields} and \eqref{Eq:ChiralFieldsh}, the result is 
\begin{align}
{\mathcal{L}^{g}}_\text{K1} 
%	&= \int d^4\theta\sqrt{g}\:{\Trp}\: 2\overline{\Phi}_1 e^{2V} {\Phi}_1\nonumber\\
	&= \sqrt{g}\Big\{
		2\overline{\phi}_1D\phi_1 + 2\overline{F}_1F_1
		-2\overline{(\nabla^m\phi_1)}(\nabla_m\phi_1) 
		- 2\overline{\phi}_1\{\sigma,\bar\sigma\}\phi_1\nonumber\\
	&\hspace{1cm}	+ 2i{\overline{\psi}_1}_{-}(\nabla_0+\nabla_1){\psi_1}_{-}
		+ 2i{\overline{\psi}_1}_{+}(\nabla_0-\nabla_1){\psi_1}_{+}\nonumber\\
	&\hspace{1cm}	-2\sqrt{2}({\overline{\psi}_1}_{+}\bar\sigma{\psi_1}_{-} 
							+ {\overline{\psi}_1}_{-}\sigma{\psi_1}_{+})
				\nonumber\\
	&\hspace{1cm} - i2\sqrt{2} ({\overline{\psi}_1}_{-}\overline{\lambda}^{-} 
							+ {\overline{\psi}_1}_{+}\overline{\lambda}^{+})\phi_1
				+ i2\sqrt{2}\: \overline{\phi}_1 ({\lambda}^{-}{{\psi}_1}_{-} + {\lambda}^{+}{{\psi}_1}_{+})\Big\}.
\end{align}

%%%%%
\subsection{Kinetic term of the chiral multiplets, $\Phi_2, \Phi_3$:  ${\mathcal{L}^{g}}_\text{K2}$}
The kinetic term of the chiral multiplets $\Phi_2$ and $\Phi_3$ is 
\begin{align}\label{Eq:LagrangianK2n}
{\mathcal{L}^{g}}_\text{K2}
	= \sum_{i=2,3}\int d^4\theta\:{\Trp}
		\left\{e^{-2V}\left(\frac12\partial_i + \overline{\Phi}_i\right)e^{2V} + \Phi_i\right\}^2.
\end{align}
The integrand of \eqref{Eq:LagrangianK2n} is invariant under the gauge transformation,
\begin{subequations}
\begin{align}
 e^{2V} 	&\rightarrow e^{-i\Lambda^\dagger}e^{2V} e^{i\Lambda},\\
 \Phi_2 	&\rightarrow e^{-i\Lambda^\dagger}\Phi_2 e^{i\Lambda}
 			 - e^{-i\Lambda}(\partial_2 e^{i\Lambda}),\\
 \Phi_3 	&\rightarrow e^{-i\Lambda^\dagger}\Phi_3 e^{i\Lambda}
 			 - e^{-i\Lambda}(\partial_3 e^{i\Lambda}).
\end{align}
\end{subequations}

The component expression is 
\begin{align}
{\mathcal{L}^{g}}_\text{K2} 
	&= {\Trp}\sum_{i=2,3} \Bigg\{
		-2D\partial_{i} \text{Re}\phi_i -4 \text{Re}\phi_i D \overline{\phi}_{i} + 2\overline{F}_iF_i\nonumber\\
	&\hspace{2cm}	-2\partial^m\text{Re}\phi_i\partial_m\text{Re}\phi_i 
				+ 4i \text{Re}\phi_i v^m\partial_m \overline{\phi}_i
				+2\text{Re}\phi_i v^\mu \partial_i v_\mu 
				-4 \text{Re}\phi_i v^{\mu} v_{\mu} \overline{\phi}_i\nonumber\\
	&\hspace{2cm}	-\frac{1}{2}\left(2\partial^m(\text{Im}\phi_i) 
				+ \partial_i v^m -2 v^m \overline{\phi}_i 
							\right)
						\left(2\partial_m(\text{Im}\phi_i) 
						+ \partial_i v_m -2 v_m \overline{\phi}_i 
							\right)\nonumber\\
	&\hspace{2cm} -\frac{1}{2}(\partial_iv^j -2 v^j\overline{\phi}_i)(\partial_iv_j -2 v_j\overline{\phi}_i)\nonumber\\
	&\hspace{2cm} +i\sqrt{2}\psi_i\left(\partial_i\lambda -2 \lambda\overline{\phi}_i 
					-i\sqrt{2}\sigma^\mu v_\mu \overline{\psi}_i 
					-\frac{1}{\sqrt{2}}\sigma^m\partial_m\overline{\psi}_i\right)\nonumber\\
	&\hspace{2cm} -\sqrt{2}\overline{\psi}_i \left(\partial_i\overline{\lambda} 
					-2\overline{\lambda}\overline{\phi}_i
					+\frac{1}{\sqrt{2}}\bar\sigma^n\partial_n\psi_i\right)
				 	+4\sqrt{2}i \text{Re}\phi_i\overline{\lambda}\overline{\psi}_i					
		\Bigg\}.
\end{align}

%%%%%%
\subsection{Potential term: ${\mathcal{L}^{g}}_{W}$}
In the non-Abelian case the potential term is modified including a commuter term:
\begin{align}
{\mathcal{L}^{g}}_{W} = 2{\Trp} \left\{\int d^2\theta\: \left(\Phi_1(\partial_2 \Phi_3 - \partial_3 \Phi_2) -2[\Phi_2,\Phi_3] \right) + \text{  c.c.  }
				\right\}.
\end{align}

We can obtain 
\begin{align}
{\mathcal{L}^{g}}_{W} 
	= 2{\Trp}\bigg\{
		F_1\epsilon^{ij}(\partial_i\phi_j -2 \phi_i\phi_j)
		+ F_i\epsilon^{ij}(\partial_j\phi_1 -2 [\phi_j,\phi_1])
		-\psi_1\epsilon^{ij}(\partial_i\psi_j -2 [\phi_i, \psi_j])
		+2 \phi_1\epsilon^{ij}\psi_i\psi_j
				\nonumber\\
	 + \qquad \text{c.c.}\:\: \bigg\}. 
\end{align}

%%%%%%
\subsection{Total Lagrangian $\mathcal{L}^{g}$}\label{Subsec:TotalNonabelLag}
Putting together the kinetic terms of the vector and chiral multiplets \eqref{Eq:LagSigma}, \eqref{Eq:LagK1}, \eqref{Eq:LagK2} and the potential term \eqref{Eq:LagW} we obtain the total Lagrangian:
\begin{align}
\mathcal{L}^{g} &:= {\mathcal{L}^{g}}_{\Sigma} + {\mathcal{L}^{g}}_\text{K1} 
	+ {\mathcal{L}^{g}}_\text{K2} + {\mathcal{L}^{g}}_{W}.
\end{align}

The bosonic term of the Lagrangian is 
\begin{align}
{\mathcal{L}^{g}}_\text{bos}
	&= \sqrt{g}\Big\{
		\frac{1}{2}D^2 + \frac{1}{2}v_{01}^2 
		+ |\nabla_0\sigma|^2 - |\nabla_1\sigma|^2 - \frac{1}{2}[\sigma,\bar\sigma]^2
		+ 2D[\phi_1,\overline{\phi}_1] + 2\overline{F}_1F_1\nonumber\\
	&\hspace{1.5cm}
		-2\overline{\nabla^m\phi_1}\nabla_m\phi_1 
		-2 \overline{\phi}_1\{\sigma,\bar\sigma\}\phi_1
		\Big\}\nonumber\\
	&\hspace{1cm}
		+\sum_{i=2,3}\Big\{ D(2[\phi_i,\overline{\phi}_i]-\partial_i2\text{Re}\phi_i)
			+2\overline{F}_iF_i
			-\frac12 \partial_iv^\mu \partial_iv_\mu
			-2\nabla^m\overline{\phi}_i\nabla_m\phi_i
			+2[v^j,\overline{\phi}_i][j_j, \phi_i]\nonumber\\
	&\hspace{2.5cm}
			-2\partial_i v^m \partial_m \text{Im}\phi_i
			+2\partial_iv^\mu [v_\mu,\overline{\phi}_i]
			+[v^\mu,\partial_iv_\mu]2\text{Re}\phi_i
			\Big\}\nonumber\\
	&\hspace{1cm}
		+2\Big\{
		F_1\epsilon^{ij}(\partial_i\phi_j - 2\phi_i\phi_j)
		+F_i\epsilon^{ij}(\partial_j\phi_1 -2[\phi_j, \phi_1])
		\Big\}.
%						\partial_{i}\text{Re}\phi_i + 2[\overline{\phi}_1,\text{Re}\phi_1]
%						\right)\nonumber\\
%	&\hspace{1cm} + \sum_{\ell = 1,2,3} 2\overline{F}_\ell F_\ell
%			 	+ \sum_{i=2,3} 2\left\{
%				F_1\epsilon^{ij}(\partial_i\phi_j - 2\phi_i\phi_j) + F_i\epsilon^{ij}(\partial_j\phi_1 - 2[\phi_j,\phi_1])
%				+ \text{c.c.}
%				\right\}\nonumber\\
%	&\hspace{1cm} +\sum_{i=2,3}\big\{
%				-2 \partial^m\text{Re}\phi_i\partial_m\text{Re}\phi_i
%				+ 4i\text{Re}\phi_iv^m\partial_m\overline{\phi}_i
%				\nonumber\\
%	&\hspace{3.5cm}
%				+ 2\text{Re}\phi_i(v^m\partial_i v_m + v^a \partial_i v_a)
%				- 4\text{Re}\phi_i (v^m v_m + v^a v_a) \overline{\phi}_i
%				\nonumber\\
%	&\hspace{3.5cm} - \frac{1}{2}\left(
%					2\partial^m(\text{Im}\phi_i) + \partial_i v^m - 2v^m\overline\phi_i
%					\right)	
%					\left(
%					2\partial_m(\text{Im}\phi_i) + \partial_i v_m - 2v_m\overline\phi_i
%					\right)\nonumber\\
%	&\hspace{9cm} -\frac{1}{2}(\partial_iv^a - 2v^a\overline{\phi}_i)(\partial_iv_a - 2v_a\overline{\phi}_i)
%	\Big\}.
\end{align}

The equations of motion of the auxiliary fields are
\begin{subequations}\label{Eq:CurvedNAeom}
\begin{align}
D &: \sqrt{g}D + \left(2\sqrt{g}[\phi_1,\overline{\phi}_1] + \sum_{i=2,3}2[\phi_i,\overline{\phi}_i] 
	-2 \sum_{i=2,3}\partial_i\text{Re}\phi_i\right) =0,\\
F_1 &: \sqrt{g}\overline{F}_1 + (\partial_2\phi_3 -\partial_3\phi_2 - 2[\phi_2,\phi_3] ) =0,\\
F_2 &: \overline{F}_2 + (\partial_3\phi_1  - 2[\phi_3,\phi_1] ) =0,\\
F_3 &: \overline{F}_3 - (\partial_2\phi_1 - 2[\phi_2,\phi_1] ) =0.
\end{align}
\end{subequations}

Eliminating the auxiliary fields and relating the components to the Yang-Mills fields in the same way as the Abelian case (see Table \ref{Table:CorrespondenceFields}), we obtain the 4-dimensional Yang-Mills Lagrangian:
\begin{align}\label{Eq:YMLagbos}
\mathcal{L}_\text{YM,bos} 
	&= \sqrt{g}\: {\Trp}\bigg\{-\frac14 F_{\mu\nu}F^{\mu\nu}
		-\frac12 \sum_{A=4,\cdots,9} \nabla'_\mu X'_A \nabla'^\mu X'_A
		+\frac14 \sum_{A, B =4,\cdots,9}[X'_A,X'_B]^2\nonumber\\
	&\hspace{2.5cm}		+\frac12 \sum_{i=2,3}\sum_{A=4,5}\partial_i\partial_ih(X'_AX'_A)
		\bigg\}\nonumber\\
	&= \sqrt{g}\: {\Trp}\bigg\{-\frac14 F_{\mu\nu}F^{\mu\nu}
		-\frac12 \sum_{A=4,\cdots,9} \nabla'_\mu X'_A \nabla'^\mu X'_A
		+\frac14 \sum_{A, B =4,\cdots,9}[X'_A,X'_B]^2\nonumber\\
	&\hspace{2.5cm}		-\frac14 R\sum_{A=4,5}(X'_AX'_A)
		\bigg\},
\end{align}
where to obtain the last form we rescaled the fields:
\begin{equation}
X'_{4,5} := \frac{1}{g^{1/4}} X_{4,5},\:\:
X'_{6,7,8,9} := X_{6,7,8,9},
\end{equation}
and used the fact that the curvature of the surface can be written as \eqref{Eq:Curvature}:
\begin{equation*}
\sqrt{g}R= - 2\sum_{i}\partial_i\partial_ih.
\end{equation*}
The covariant derivative is changed by the twist:
\begin{subequations}
\begin{align}
\nabla'_{i} X'_{4} &= \partial_{i}X'_{4} + i[A_{i}, X'_{4}] -\mathcal{A}^{45}_i X'_{5} ,\\
\nabla'_{i} X'_{5} &= \partial_{i}X'_{5} + i[A_{i}, X'_{5}] -\mathcal{A}^{54}_i X'_{4} ,\\
\nabla'_{i} X'_{\ell} &= \partial_{i}X'_{\ell} + i[A_{i}, X'_{\ell}]\:\: ;\:\: \ell = 6,7,8,9,
\end{align}
\end{subequations}
where the external field $\mathcal{A}^{45}_i = -\mathcal{A}^{54}_i$ is defined as $\mathcal{A}^{45}_i = - \Omega^{23}_i$.
Therefore, we obtain the result that the superfield Lagrangian \eqref{Eq:YMLagbos} is equivalent to the 4-dimensional Yang-Mills Lagrangian on the Riemann surface, which is the bosonic part of the Lagrangian \eqref{Eq:YMlagrangian} after twisting.

%%%%%%
\section{Discussion}\label{Sec:Discussion}
Our aim was to construct the 4-dimensional super Yang-Mills action in terns of 2-dimensional superfields. 
Redefining the inner product of Lie algebra including the integral on the inner space $\Sigma_\mathbf{g}$,
\begin{equation}
\int_{\Sigma_\mathbf{g}} dx^2dx^3 {\Trp} =: \hat{\Trp},
\end{equation}
we can interpret the action \eqref{Eq:NALagrangian} as defined for 2-dimensional fields:
\begin{align}
\int d^4x\: \mathcal{L}^{g}
		&= \int_{R^{1,1}} d^2x \hat{\Trp}\sqrt{g}\: \bigg[ - \int d^4\theta\: \overline{\Sigma}\Sigma 
			+ 2 \int d^4\theta\: e^{-2V}\overline{\Phi}_1e^{2V}\Phi_1
			\bigg]\nonumber\\
		&\hspace{1cm}	+ \int_{R^{1,1}} d^2x\: \hat{\Trp}\int d^4\theta \sum_{i=2,3} \bigg\{e^{-2V}\left(\frac12\partial_i +\overline{\Phi}_i\right)e^{2V} + \Phi_i\bigg\}^2
				\nonumber\\
		&\hspace{1cm}	+ 2\int_{R^{1,1}} d^2x\: \hat{\Trp} \bigg[ 
		 \left(\int d^2\theta\: \Phi_1(\partial_2 \Phi_3 - \partial_3 \Phi_2)
					-2[\Phi_2,\Phi_3] \right)+ \text{  c.c.  }
				\bigg].
\end{align}

In the construction of the Lagrangian in sections \ref{Sec:Abelian} and \ref{Sec:NAbelian}, we treated only the bosonic part of the Lagrangian for simplicity.
However, for the curved space we treated in the previous section, the fermonic part of the Lagrangian,
\begin{align}\label{Eq:YMLagfer}
\mathcal{L}_\text{YM,fer} = \frac{1}{(g_\text{YM})^2}
		\sqrt{g}\: {\Trp}
		\Big\{\frac{i}{2}\overline{\Psi}\Gamma^\mu \nabla'_\mu \Psi
			-\frac{1}{2}\overline{\Psi}\Gamma^A[X_A',\Psi]
			\Big\},
\end{align}
should be obtained, as expected by the supersymmetry, from the terms in \eqref{Eq:NALagrangian} including the fermions, $\psi_1,\psi_2,\psi_3$ and $\lambda$.

As we can see in \eqref{Eq:YMLagbos}, the mass term of scalar fields $X_A'$ exist.
Indeed, this term is needed to possess the supersymmetry. 

One of interesting future work is to find a counterpart of a triple $(\rho, H, \mathfrak{B})$ which characterizes the boundary condition on $R^{1,3}$ \cite{Gaiotto:2008sa, Gaiotto:2008ak} and to analyze S-duality of these supersymmetric boundary conditions. 

%%%%%%%
\section*{Acknowledgement}
I would like to thank Satoshi Yamaguchi, Satoshi Iso and Tadashi Okazaki for useful discussions and comments.

%%%%%%%
\begin{appendix}
\section{Notation}\label{Sec:Notation}
We summarize here the notation used in this paper.
Our notation is the same as Witten, \cite{Witten:1993yc}.
The 4d SYM is defined on $R^{1,1}\times \Sigma_\mathbf{g}$.
$R^{1,1}$ is coordinated by $x^m,\: m=0,1$ and the Riemann surface $\Sigma_\mathbf{g}$ is coordinated by $x^i,\: i=2,3$. The metrics is 
\begin{align}
g_{mn}=\text{diag}(-1,+1), \:\:
g_{ij} = e^{2h(x^2,x^3)}\delta_{ij},
\end{align} 
where $h(x^2,x^3)$ is a function of $x^2$ and $x^3$.

We use the following differential operators to define the supersymmetry transformation:
\begin{align}
{Q}_{\alpha} := \frac{\partial}{\partial \theta^{\alpha}} 
	- i(\sigma^{\mu}\bar\theta)_{\alpha}\frac{\partial}{\partial x^{\mu}},\:\:
\overline{{Q}}_{\dot\alpha} := -\frac{\partial}{\partial\bar\theta^{\dot\alpha}}
	+ i(\theta\sigma^{\mu})_{\dot\alpha}\frac{\partial}{\partial x^{\mu}}.
\end{align}
The supersymmetry transformation is $\delta = \xi Q + \bar\xi \bar{Q}$.
We also define the super-covariant derivatives:
\begin{align}
{D}_{\alpha} := \frac{\partial}{\partial \theta^{\alpha}} 
	+ i(\sigma^{\mu}\bar\theta)_{\alpha}\frac{\partial}{\partial x^{\mu}},\:\:
\overline{D}_{\dot\alpha} := -\frac{\partial}{\partial\bar\theta^{\dot\alpha}}
	- i(\theta\sigma^{\mu})_{\dot\alpha}\frac{\partial}{\partial x^{\mu}}.
\end{align}
We use an unusual ordering of sigma matrices:
\begin{align}
\sigma^\mu = \left\{
	\sigma^{m}, \sigma^{i}
	\right\}
	= \left\{
\left(\begin{array}{cc}-1 & 0 \\0 & -1\end{array}\right),
\left(\begin{array}{cc}1 & 0 \\0 & -1\end{array}\right),
\left(\begin{array}{cc}0 & 1 \\1 & 0\end{array}\right),
\left(\begin{array}{cc}0 & -i \\i & 0\end{array}\right)
\right\}.
\end{align}
We use the notation for integration for Grassmann coordinates:
\begin{align}
\int d^4\theta\: F &= F \Big|_{\theta\theta\bar\theta\bar\theta} 
				= \frac{1}{4} F\Big|_{\theta^{-}\theta^{+}\bar\theta^{+}\bar\theta^{-}},\\
\int d^2\theta\: F &= F \Big|_{\theta\theta} 
				= \frac{1}{2} F\Big|_{\theta^{-}\theta^{+}}.
\end{align}

\end{appendix}
%%%%%%
%\bibliographystyle{utphys} % Choose Phys. Rev. style for bibliography
%\bibliography{myrefs}
\providecommand{\href}[2]{#2}\begingroup\raggedright\endgroup

\end{document}